\documentclass[a4paper]{article}
\usepackage{epsfig}
\usepackage{color}

\newcounter{comnum}
\begin{document}
\begin{flushright}
LAPTH--1141/06
\end{flushright}
\vskip 0.2cm

\centerline{\bfseries \large Is the dark matter interpretation of the EGRET gamma excess}
\centerline{\bfseries \large compatible with antiproton measurements?}

\bigskip

\centerline{Lars Bergstr\"om, Joakim Edsj\"o, Michael Gustafsson}
\smallskip
\centerline{\em Department of Physics, Stockholm University, AlbaNova University Center,}
\centerline{\em SE-106 91 Stockholm, Sweden}

\bigskip

\centerline{Pierre Salati}
\smallskip
\centerline{\em Laboratoire d'Annecy-le-Vieux de Physique Th\'eorique LAPTH, CNRS and Universit\'e de Savoie,}
\centerline{\em 9, Chemin de Bellevue, B.P.110 74941 Annecy-le-Vieux, France}

\begin{abstract}
We investigate the internal consistency of the halo dark matter
model which has been proposed by de Boer et al.\ to explain the
excess of diffuse galactic gamma rays observed by the EGRET
experiment.
Any model based on dark matter annihilation into quark jets, such as
the supersymmetric model proposed by de Boer et al., inevitably also
predicts a primary flux of antiprotons from the same jets. Since
propagation of the antiprotons in the unconventional, disk-dominated
type of halo model used by de Boer et al.\ is strongly constrained
by the measured ratio of boron to carbon nuclei in cosmic rays, we
investigate the viability of the model using the DarkSUSY package to
compute the gamma-ray and antiproton fluxes.
We are able to show that their model is excluded by a wide
margin from the measured flux of antiprotons.
We therefore find that a model of the type suggested by Moskalenko
et al., where the intensities of  protons and electrons in the
cosmic rays vary with galactic position, is far more plausible to
explain the gamma excess.
\end{abstract}


\section{Introduction}

In a series of papers, de Boer et al.\ \cite{boer-short, boer-long,
deBoer:2004es, deBoer:2003ky} have put forward the idea that the
well-known EGRET excess of diffuse galactic gamma rays
\cite{egret_excess} could be well explained by dark matter
annihilations. The idea that excess from the region near the
galactic center may be due to dark matter annihilations has a long
history (at least \cite{BUB,Gondolo:1998dg,Cesarini:2003nr}) but de
Boer et al.~have extended this idea to claim that all the diffuse
galactic gamma rays detected above 1 GeV by the EGRET satellite,
irrespective of the direction, has a sizeable dark matter
contribution (for a similar, but less explicit, proposal, see
\cite{dixon}). De Boer et al.~propose specific supersymmetric models
as examples of viable models with correct relic density, and the
gamma-ray fluxes are made to fit the observations. The price they
have to pay, however, is a rather peculiar dark matter halo of the
Milky Way, containing massive, disc concentrated rings of dark
matter besides the customary smooth halo. In addition, they have to
crank up the predicted gamma-ray flux by considerable ``boost
factors". We will here examine these hypotheses a bit closer. In
particular, we will discuss the astrophysical properties and
possible inconsistencies of the model. We point out that, besides
the strange features of the halo model (that we judge difficult to
achieve with non-dissipative dark matter), supersymmetric models
with large gamma-ray flux are essentially always accompanied by a
large antiproton flux (see, e.g., \cite{Rudaz:1987ry,BEU}). We
investigate what the antiproton fluxes would be in the same halo
model, using model-by-model the same boost factors as needed to fit
the gamma-ray spectrum. We find that low-mass models (masses less
than 100 GeV) that have low boost factors tend to overproduce
antiprotons by a factor of around ten. Higher-mass models (above a
few hundred GeV) have a lower antiproton rate, so the overproduction
is slightly less. However, they give hardly any improvements to the
fits to the gamma-ray spectrum. We will perform this analysis in a
general Minimal Supersymmetric Standard Model (MSSM), but as the
correlation between gamma rays and antiprotons is a general feature,
our results will be more general. Our conclusion is that the
proposal of de Boer et al.\ \cite{boer-long} to explain the gamma
excess in all sky directions is, at present, not viable\footnote{We
note that earlier versions of the scenario of de Boer et al., e.g.
\protect\cite{deBoer:2003ky}, which had higher mass models favored,
did include a discussion of the antiproton fluxes. In the later
papers with lower-mass models, they do not discuss the issue. We
furthermore conclude that in the earlier papers, where a NFW
\cite{NFW} dark matter profile was used, the EGRET observations in
\emph{all} sky directions were not explained and, consequently, dark
matter annihilations could not alone explain all the gamma excess.}.
Although -- of course -- we cannot exclude a small contribution to
the diffuse gamma-ray flux from dark matter annihilations.

\section{Description of the model} \label{sec:desc}

\begin{figure}[t]
\centerline{
\epsfig{file=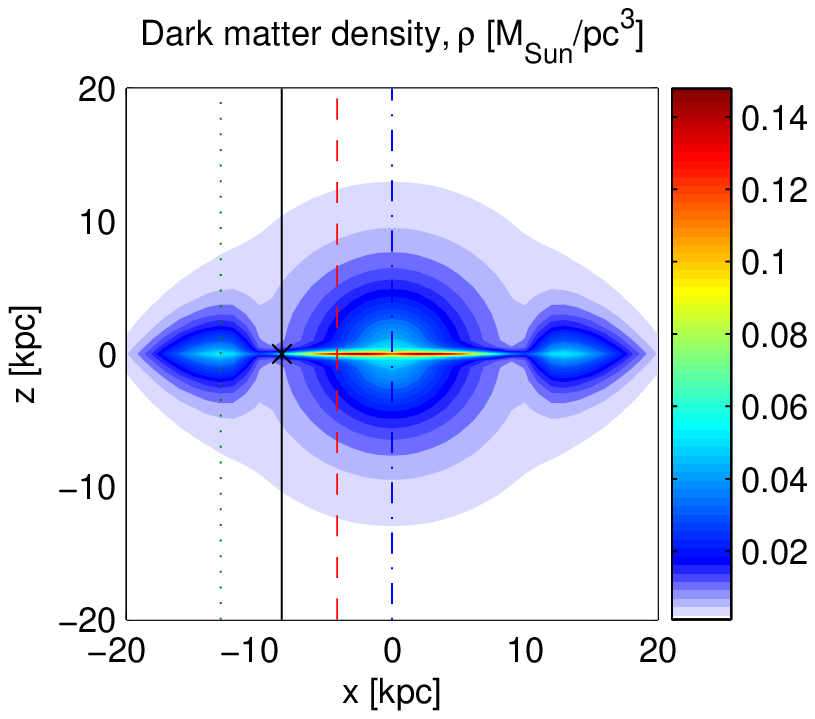,width=0.48\textwidth}
\epsfig{file=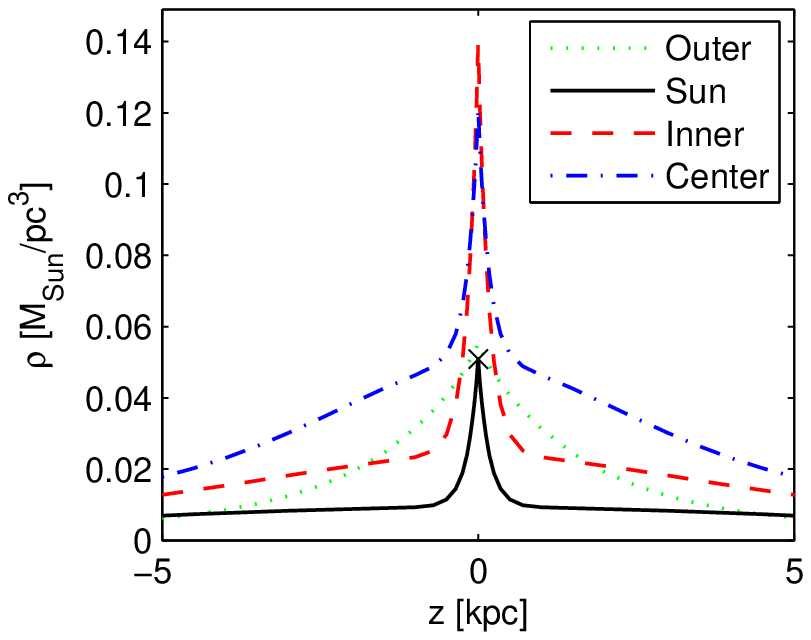,width=0.48\textwidth}}
\centerline{
\epsfig{file=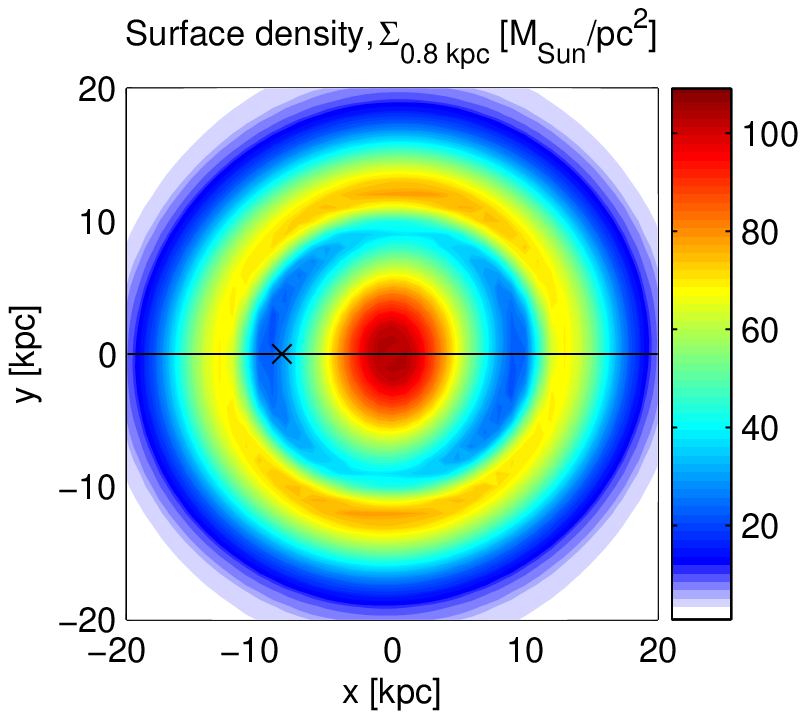,width=0.48\textwidth}
\epsfig{file=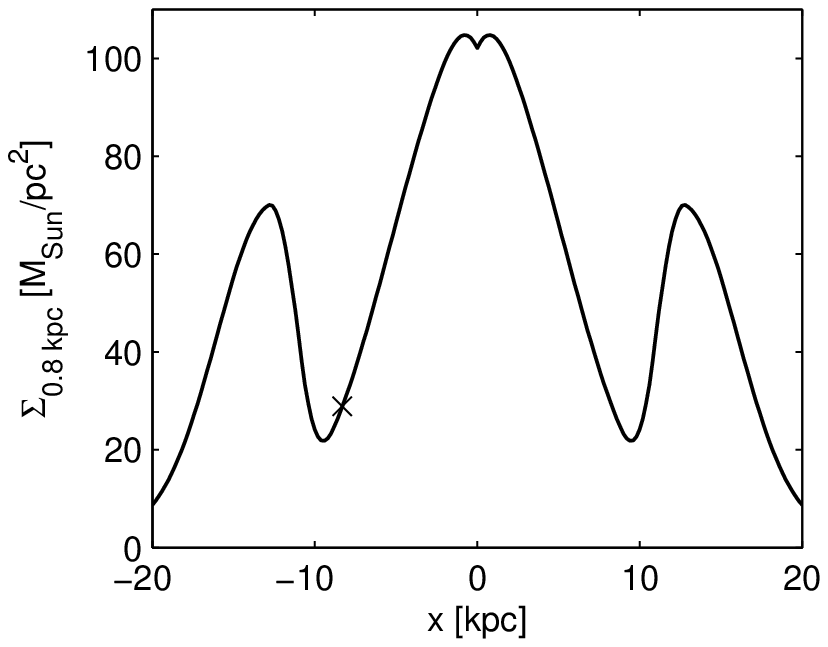,width=0.48\textwidth}}
\caption{The dark matter distribution in the halo model of de Boer
et al. \cite{boer-long}. The upper panel shows the concentration of
dark matter to the galactic disc, where in the right figure the
density dependence is explicitly plotted versus the vertical
distance from the galactic plane -- at the position of the outer
ring (dotted/green), solar system (solid/black), inner ring
(dashed/red) and galactic center (dash-dotted/blue). The lower panel
shows the dark matter surface mass density within 0.8 kpc from the
galactic disc. The Earth's location is shown with a $\times$-sign.}
\label{fig:deboer-density}
\end{figure}

Gamma rays have the advantage of pointing back directly to their
sources in the Galaxy and not to suffer from energy losses. This,
together with known gamma-ray spectral shape from dark matter
annihilation (distinct from the conventional background), permit to
extract the sky-projected dark matter distribution from the EGRET
observations. Taking this one step further de Boer et al. propose a
specific dark matter profile, with 18 free parameters, and do a best
fit to the EGRET data (for details, see \cite{boer-long}). The
density profile de Boer et al.\ obtain consists of a dark matter
halo with the following ingredients:
\begin{itemize}
   \item a triaxial smooth halo,
   \item an inner ring at about 4.15 kpc with a density falling off as
$\rho \sim e^{-|z|/\sigma_{z,1}} \; ; \; \sigma_{z,1} = 0.17
\mbox{~kpc}$, and
   \item an outer ring at about 12.9 kpc with a density falling off as
$\rho \sim e^{-|z|/\sigma_{z,2}} \; ; \; \sigma_{z,2} = 1.7
\mbox{~kpc}$.
\end{itemize}
where $z$ is the height above the galactic plane.

The triaxial halo is a modified isothermal sphere, but flattened in
the direction of the Earth and in the $z$-direction. The inner ring
is rather broad, but very closely located to the baryonic disc, with
an exponential fall-off above the galactic plane. The outer ring is
more massive and slightly narrower and also exhibits an exponential
fall-off above the galactic plane. The outer ring is also taken to
have a sharper fall-off on the inside than the outside. Both rings
are elliptic. The details of the parameterization of the dark matter
model can be found in \cite{boer-long}. In
Fig.~\ref{fig:deboer-density}, we show the strong concentration of
dark matter to the disc (upper panel) as well as the ring structure
of the model (lower panel). The steep exponential density fall-off
away from the disc, originating from the two rings, can most clearly
be seen in the upper-right plot.

Since conventional models of the diffuse gamma-ray emission employ
scattering of cosmic rays on gas and dust in the galactic disc, we
may already note that this model will produce a gamma-ray flux that
has an angular distribution very similar to the generally accepted
cosmic ray contribution \cite{mosk-strong-reimer}. In fact, besides
the need for measurements with a better energy resolution such as
will be given by GLAST, the desired spectral shape can also be
obtained by adjusting the balance between the contributions from
proton and electron cosmic rays (whose intensity is very poorly
constrained away from the solar neighborhood)
\cite{mosk-strong-reimer}. In \cite{mosk-strong-reimer} it was shown
that one can get a good agreement with EGRET data by adjusting the
electron and proton injection spectra (without any need for a dark
matter contribution).

\section{Astrophysical problems with the model}

Even though the dark matter halo profile by de Boer et al.\ explains
the EGRET data very well, we will here go through some of the
astrophysical objections to this model. First, one may notice that
the model of the dark matter really is very close to the
``standard'' model for the baryons of the Milky Way, containing a
thin and a thick disc and a central bulge (see, e.g., \cite{sdss}).
Since the dark halo is much more massive than the baryonic one, one
of the first things one should investigate is whether there is room
to place as much unseen matter in the vicinity of the disc as de
Boer et al.\ do.

\subsection{Disc surface mass density}

By observations of the dynamics and density fall-off of stars in the
disc, one can get a measure of the gravitational pull perpendicular
to the galactic plane. This in turn can be converted to an allowed
disc surface mass density, a method pioneered by Bahcall
\cite{bahcall}. Recent analyses \cite{Korchagin,HolmbergFlynn04,
Bienayme05} of the disc surface mass density at the solar system
location  have converged to a model with little room for a
concentration of dark matter in the disc. Observations are well
described by a smooth dark matter halo and a disc of identified
matter (mainly containing stars, white and brown dwarfs and
interstellar matter in form of cold and hot gases).
%
\begin{table}
\begin{center}
{\begin{tabular}{@{}cr@{ -- }lcr@{ -- }lcc@{}} \hline \hline
surface density: &   \multicolumn{2}{l}{dynamical }      &  identified      & \multicolumn{2}{c}{unidentified}   &  dark matter in \cite{boer-long}\\
                 &   \multicolumn{2}{c}{$M_\odot/pc^2$ } &  $M_\odot/pc^2$  & \multicolumn{2}{c}{$M_\odot/pc^2$} &  $M_\odot/pc^2$ \\
\hline
$\Sigma_{\rm 50\,pc}$   & \quad9 & 11  & $\sim$  9 &\quad\,0 &  2  & 4.5 \\
$\Sigma_{\rm 350\,pc}$  &  36    & 48  & $\sim$ 34 &   2      & 14  & 19  \\
$\Sigma_{\rm 800\,pc}$  &  59    & 71  & $\sim$ 46 &  13      & 25  & 29  \\
$\Sigma_{\rm 1100\,pc}$ &  58    & 80  & $\sim$ 49 &   9      & 32  & 35  \\
\hline \hline
\end{tabular}}
\end{center}
\caption{Measured local surface densities $\Sigma_{\rm |z|}$, within
heights $|z|$, compared to the amount of dark matter in the model of
de Boer et al.~\cite{boer-long}. The amount of dark matter exceeds
the allowed span for unidentified gravitational matter  in the inner
part of the galactic disc (i.e.\ around $z=0$).
\cite{HolmbergFlynn98,HolmbergFlynn04}
 \label{tab:density}}
\end{table}
%
Table~\ref{tab:density} shows the observed local surface mass
density in both identified components and the total dynamical mass
within several heights. Their differences then give an estimate of
the allowed amount of dark matter in the local disc\footnote{Even
though the estimates are done under the assumption of a smooth dark
matter halo these constraint still holds even for a wide range of
varying dark matter profiles, including the profile concerned here
(private communication, C.~Flynn and J.~Holmberg)} -- and the result
is an exclusion of such strong concentration of unidentified/dark
matter as derived in \cite{boer-long}. All these estimates also
agree well with the observed local density of $0.102 \pm 0.006
M_\odot$/pc$^3$ in dynamical mass and $0.092 M_\odot$/pc$^3$ in
visible matter \cite{HolmbergFlynn98,HolmbergFlynn04}. This gives
room for only about $0.01M_\odot$/pc$^3$ in unidentified matter,
which should be compared to the dark matter density of 0.05
$M_\odot$/pc$^3$ in the model of de Boer et al. \cite{boer-long}.

From these numbers alone, the de Boer model gives a too high mass
density. One should keep in mind, however, that the estimates of the
possible amount of dark matter are somewhat uncertain and that the
disc models also have uncertainties of the order of 10\% in their
star plus dwarf components and uncertainties as large as about 30\%
in their gas components. Also the de Boer et al.\ model could easily
be modified to give a lower disc surface mass density at the solar
system's location in the Galaxy. However, such a modification just
to circumvent this problem seems fine-tuned. One should keep in mind
that the model, as it now stands, already have made these
modifications, e.g.\ the outer ring is Gaussian in shape in the
radial direction, except that it is forced to zero relatively fast
on the inside. This {\em ad hoc} cut-off lets the outer ring be very
massive, while keeping the local density unaffected. In
Fig.~\ref{fig:deboer-density}, the disc surface mass density for the
halo model of de Boer et al.\ is shown. We clearly see that in this
model, the Earth is located in a region with relatively low disc
mass surface density. Modifying the model to reduce this further
would make the model even more fine-tuned. It is also not unlikely
that, e.g.\, recent SDSS data \cite{sdss} could put constraints on
the halo model further away from the solar system, but such an
analysis has, to our knowledge, not been performed yet.

To summarize, the de Boer model seems to be in slight conflict with
the estimates of the disc surface mass density. It is certainly
possible to modify their dark matter model to avoid exceeding the
allowed local disc surface mass density, but the model would become
even more fine-tuned with such a modification. In fact, the proposal
by de Boer et al.\ to explain the gamma excess by introducing dark
matter and avoid any fine-tuning of the electron and proton
injection spectra as in \cite{mosk-strong-reimer}, actually ends up
being considerably fine-tuned itself.

\subsection{Ring profiles}

The rings have an exponential fall-off above the galactic plane as
given in Sec.~\ref{sec:desc}. Actually, the outer ring is claimed
\cite{boer-long} to be consistent with an observed overdensity of
stars, the Monoceros stream \cite{newberg,rocha-pinto}. However, a
recent analysis of Sloan Digital Sky Survey (SDSS) data \cite{sdss}
indicates that this is not really a ring but rather a localized
structure like that expected from a merging dwarf galaxy with tidal
arms, as already conjectured in \cite{rocha-pinto}. In the original
EGRET paper, this structure is clearly visible, but not as a ring
(as it is not visible in quadrant IV in the notation of
\cite{egret_excess}, Fig.~9).

In most models of cold dark matter, one would expect the dark matter
distribution to be much more isotropic than that of the baryonic
disc material, which supposedly forms dissipatively with energy loss
but very little angular momentum loss \cite{fall}. In fact, we know
of no reasonable example, from numerical simulations, that would
indicate that the model shown in Fig.~\ref{fig:deboer-density}
resembles a dark matter halo.

\section{Comparison with antiproton data}

%
\begin{figure}[t]
\centerline{\epsfig{file=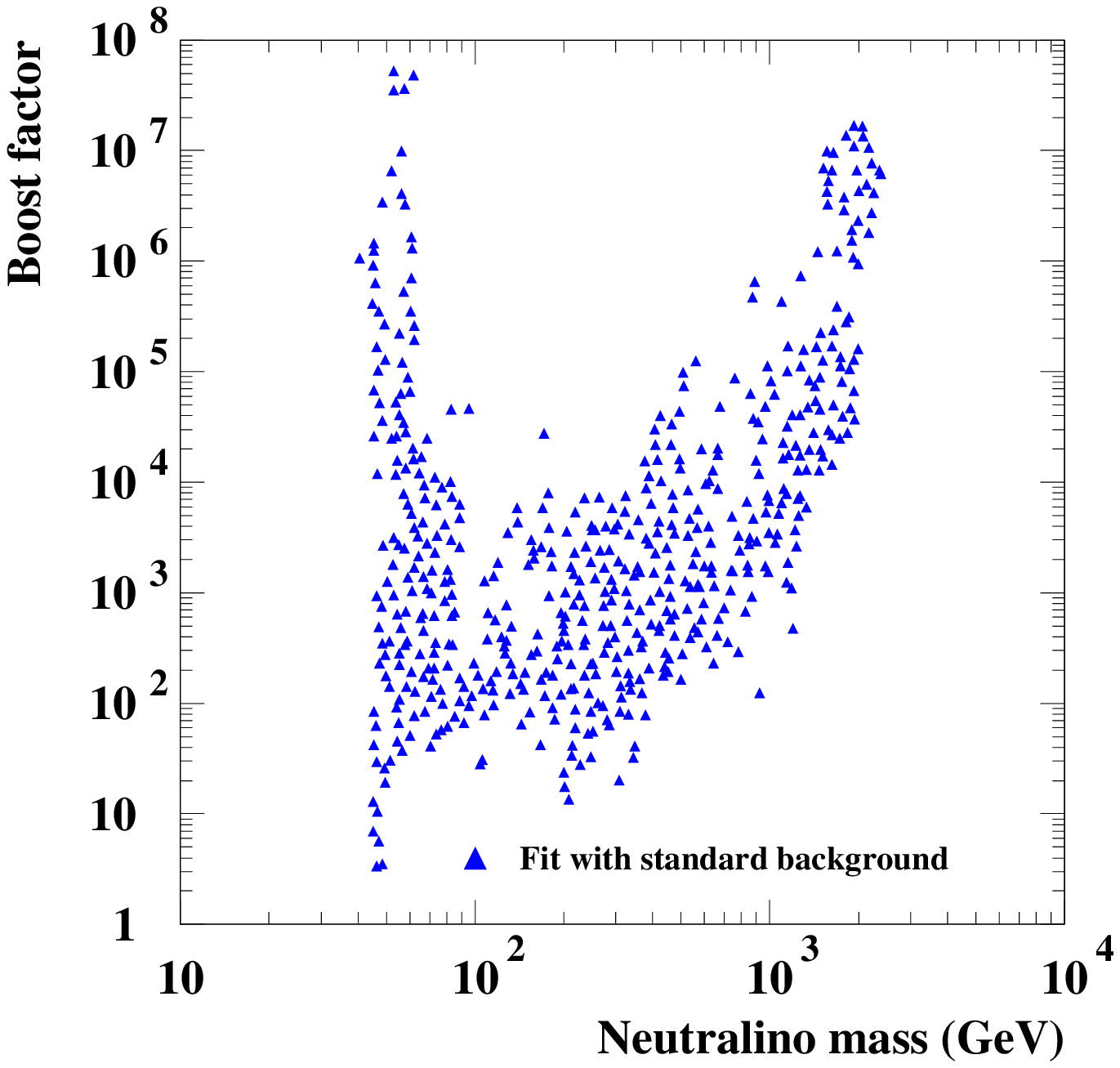,width=0.49\textwidth}
\epsfig{file=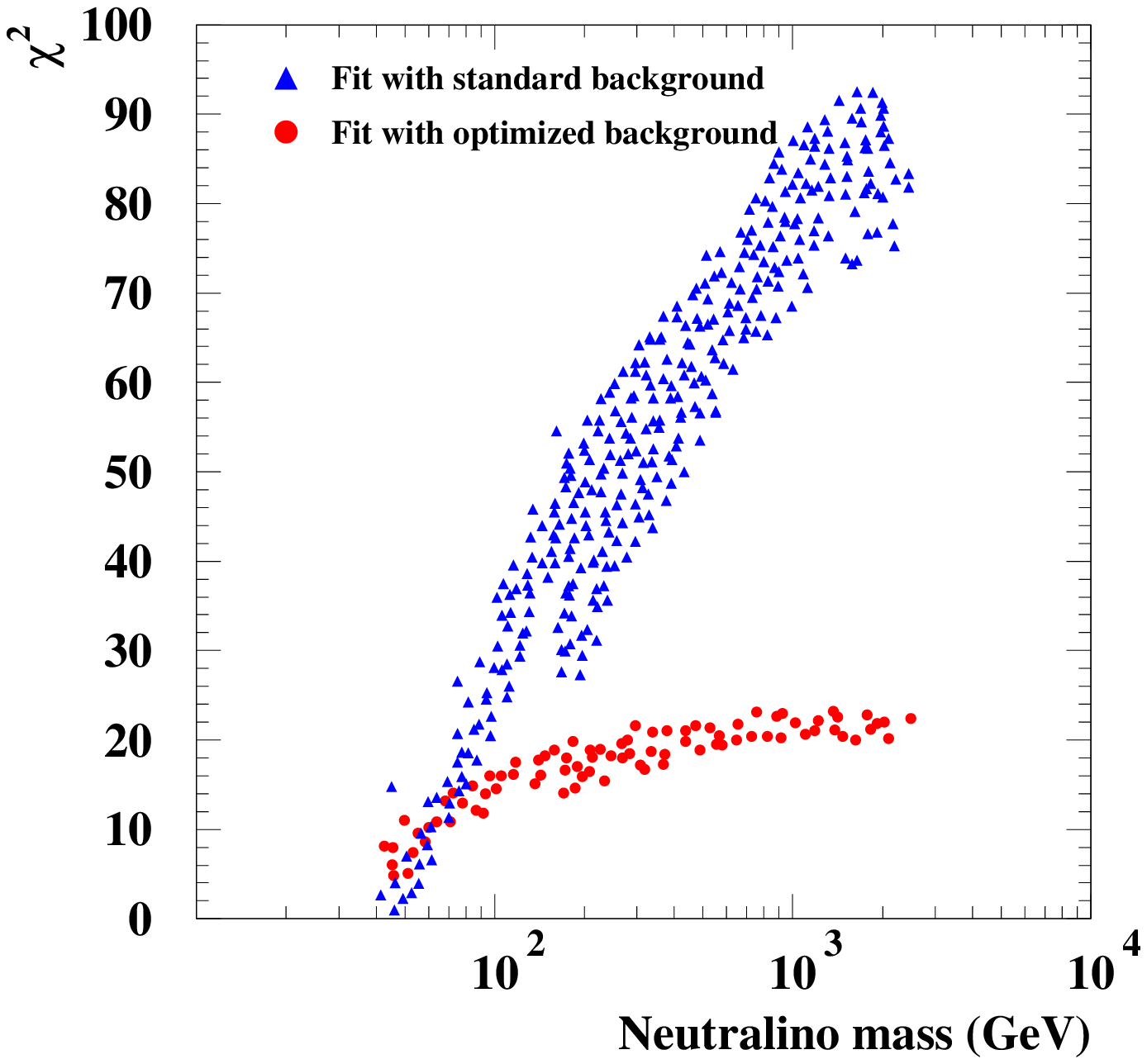,width=0.49\textwidth}}
\caption{a) The boost factor needed for a good fit of background and
a dark matter signal to the EGRET data in region A. b) The $\chi^2$
of the fits (8-2=6 degrees of freedom).} \label{fig:deboer-boost}
\end{figure}
%

\subsection{Calculating the antiproton fluxes}

We will here investigate what antiproton fluxes, from annihilation
in the dark matter halo, that the de Boer et al.\ model would
predict. To do this, we follow the approach by de Boer et al.\
\cite{boer-long} as closely as possible and proceed as follows:

\begin{enumerate}

\item We fit the expected background and signal to EGRET's observed
diffuse gamma flux towards the galactic center, with the
normalization of the background\footnote{We note that in the
analysis of de Boer et al.\ they also subtract data in regions near
the center and the disc (see Fig.3 in \cite{boer-long}) and then put
it back as signal with similar angular distribution. The only
difference is the spectral shape, which is claimed to fit better a
contribution from dark matter annihilation.} and signal as free
parameters. We here use region A in \cite{boer-long}; which is
towards the galactic center ($\pm 30^\circ$ in longitude and $\pm
5^\circ$ in latitude from the galactic center).
\item From the best-fit signal normalization, we derive the
required boost factor for the gamma flux from dark matter
annihilation .
\item Following de Boer et al.\, we assume that this boost factor
is independent of location in the Galaxy (see comment in Sec.
\ref{sec:pbar-uncert2} below though).
\item As the boost factor is assumed to be independent of location in
the Galaxy, the same boost factor would apply also for the
antiproton fluxes. We thus calculate the expected antiproton flux
for the de Boer et al.\ halo profile and then boost it by the same
boost factor as found from the EGRET fit.
\end{enumerate}

For the background gamma flux, we have used both the standard
background (as given in Fig.~2 of \cite{boer-long}) and the
optimized background (as given in Fig.~6 of \cite{boer-long}). The
difference between these are that the optimized background have a
modified proton and electron injection spectrum, trying to give a
good description of the EGRET data without dark matter. Both these
models are based upon models by Strong, Moskalenko and Reimer
\cite{mosk-strong-reimer}.

The differential gamma-ray flux in a given direction is given by
\begin{equation}
  \Phi_\gamma = 9.35 \times 10^{-10}
   \left(\frac{dN_\gamma}{dE_\gamma}\frac{\sigma
v}{10^{-29} {\rm cm}^3 {\rm s}^{-1}} \right)
   \left(\frac{{\rm GeV}}{m_\chi}
\right)^2 J(\theta,\phi) \mbox{~ cm$^{-2}$ s$^{-1}$ sr$^{-1}$
GeV$^{-1}$}
\end{equation}
with
\begin{equation}
  J(\theta,\phi) = \frac{1}{8.5~\rm kpc} \int_{\rm line~of~sight} \left(
\frac{\rho(\theta,\phi,l)}{0.3~{\rm GeV}\,{\rm cm}^{-3}} \right)^2 dl
\end{equation}
Averaged over region A towards the galactic center, we find that
$\langle J \rangle = 123.3$.

For the signal, we use DarkSUSY \cite{darksusy} to calculate the
dark matter annihilation cross sections and gamma-ray yields. We
then use the Eqs.\ above to calculate the flux in region A.
Following de Boer et al., we calculate the flux in 8 EGRET bins from
0.07 GeV to 10 GeV. We have done this for a set of rather general
MSSM models, where we have varied the standard MSSM parameters
($\mu$, $M_2$, $m_0$, $\tan \beta$, $A_b$, $A_t$ and $m_A$) between
very generous bounds (up to several TeV for the mass parameters). We
have only kept models that do not violate any accelerator
constraints and that has a relic density in the WMAP preferred range
$0.104 \le \Omega h^2 \le 0.121$. We have about 100\,000 models
satisfying these constraints (but make the plots with binned results
as the density of points has no physical meaning).

In Fig.~\ref{fig:deboer-boost}a we show the boost factors of the
signal that gives the best fits. We see that we roughly agree with
the results in \cite{boer-short} that the required boost factors are
of order 10 or more. We have also calculated the boost factors
needed when we use the so called optimized background in our fits.
We here also get roughly the same result as in \cite{boer-long},
i.e.\ the required boost factors are about a factor of 3--4 lower.

In Fig.~\ref{fig:deboer-boost}b, we show the $\chi^2$ of the fits
(divide by 8-2=6 to get the reduced $\chi^2$). As already shown by
de Boer, it is possible to get very good fits for masses below about
100 GeV. We also note that the fits with the optimized background
are never as good as with the standard background. On the other
hand, the fits are not as bad for higher masses though, as there is
less need for a signal in this model. One should note that we use
relative errors of only 7\% for the gamma fluxes, although the
overall uncertainty is often quoted to be 10-15\%
\cite{esposito,mosk-strong-reimer}. The reason for using smaller
errors is to take into account the fact that the relative
point-to-point errors are expected to be less than the overall
normalization error; dominated by systematics. The overall
normalization uncertainty is irrelevant here, since in the fitting
the normalization is a free parameter anyway. We follow de Boer et
al. and use their estimate of 7\% for the relative errors in our
$\chi^2$-fits. We note that our results are not sensitive to this
choice though (apart from the actual $\chi^2$ values of course).

\subsection{Comparing with BESS data}

We are then ready to compare with antiproton measurements. We use
DarkSUSY \cite{darksusy} to calculate the antiproton fluxes for each
of these MSSM models and with the halo profile of de Boer et al. As
our calculation assumes an axisymmetric halo profile, we symmetrized
the de Boer et al profile in such a way that we for any given radius
$r$, take an average of the source function ($\sim \rho^2$) over the
azimuthal angles. We do not expect this to introduce any large
errors as the antiproton diffusion effectively smears out local
variations in the density. In fact, from the variations of the
density over the azimuthal angles, we don't expect this
symmetrization to introduce larger errors than at most a factor of
1.5.

We have chosen to primarily compare with BESS (Balloon-borne
Experiment with a Superconducting Spectrometer) data from 1998
\cite{BESS98}, where we have chosen the bin at 0.40--0.56 GeV\@. The
reason we choose the BESS 98 data is that for these data the solar
modulation parameter is relatively low ($\phi_F=610$ MV) and the
reason we choose a low-energy bin is that here the signal is
expected to be relatively high compared to the background.

Applying the same boost factors for the antiproton fluxes as
obtained from the fit to EGRET data we show in
Fig.~\ref{fig:deboer-pbar}a the expected antiproton flux from both
the fit with the standard background and the fit with the optimized
background. The models with low masses, that have low $\chi^2$
clearly overproduces antiprotons by a large amount. In
Fig.~\ref{fig:deboer-pbar}b we show (for the standard background)
the antiproton flux, but coded with $\chi^2$. We have here also
chosen to impose a cut on the boost factor, to only allow models
with reasonably low boost factors. To be conservative, we have
allowed the boost factor to be as high as 100, which is higher than
expected from recent analyses (see e.g.~\cite{berezinsky,pieri}). It
is fairly evident that all the models with good fits to the EGRET
data give far too high antiproton fluxes.

We have also compared with a set of mSUGRA rates (the same ones used
in \cite{msugra-indirect}). The mSUGRA models show the same general
pattern as the MSSM models (i.e. similar boost factors needed,
similar best-fit $\chi^2$ etc.) so we do not show these figures
here. The model proposed by de Boer et al.\ in \cite{boer-short}
also has the same phenomenological properties as our MSSM models.
Hence, our MSSM models should be representative as a fair sample of
what you could expect for supersymmetric neutralino dark matter. In
fact, since antiprotons and gamma rays are so strongly correlated in
general, our results should be valid for more general WIMPs as well.

Other dark matter candidates, like Kaluza Klein (KK) dark matter,
would also give a similar behavior since the gamma rays and
antiprotons are so correlated. However, for e.g. KK dark matter from
Universal Extra Dimensions \cite{UED} we would not improve the fits
to EGRET data as only heavier models are favored by the relic
density constraint.

\subsection{Antiproton propagation uncertainties}
\label{sec:pbar-uncert}

\begin{figure}[t]
\centerline{\epsfig{file=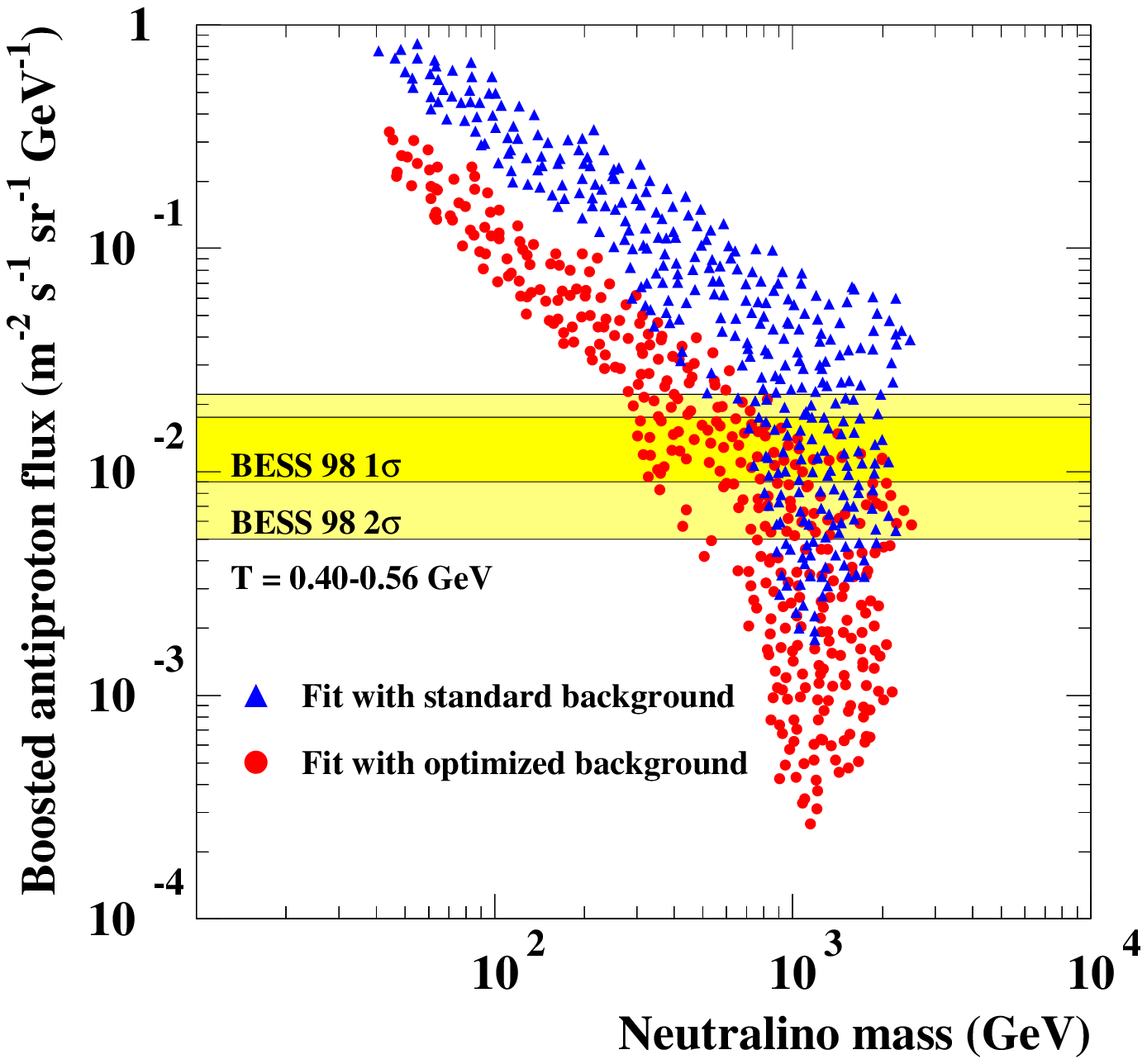,width=0.49\textwidth}
\epsfig{file=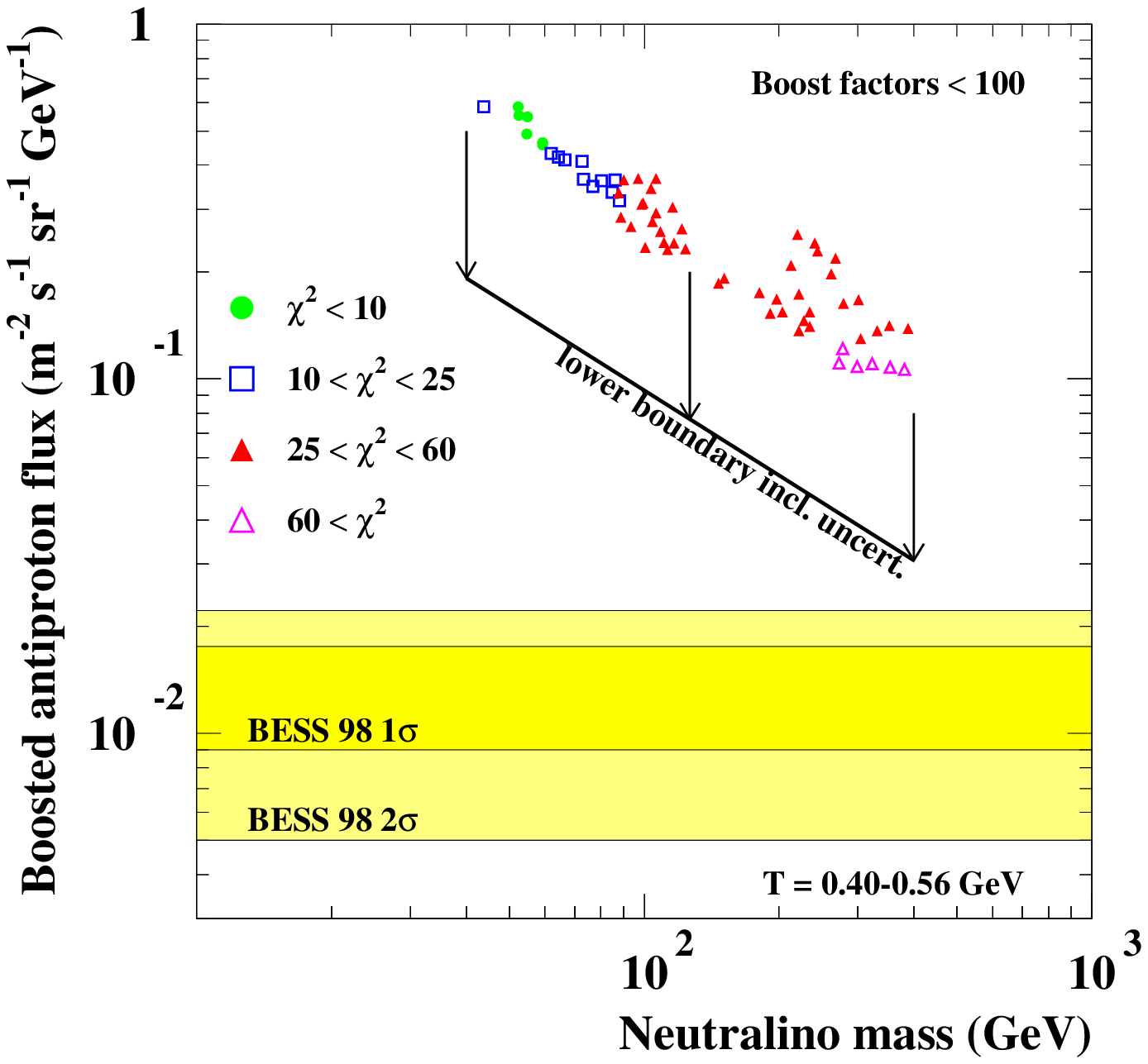,width=0.49\textwidth}}
\caption{The antiproton fluxes boosted with the same boost factor as
found for the gamma rays. In the left panel, the antiproton fluxes
for both standard and optimized background is shown, whereas in the
right panel, the $\chi^2$ of the fits has been indicated (8-2=6
d.o.f.). In the right panel, we also indicate with a solid line how
far down we could shift the models by choosing an extreme minimal
propagation model (see section \ref{sec:pbar-uncert}).}
\label{fig:deboer-pbar}
\end{figure}

Of course, it is a well-known fact that the antiproton flux from
dark matter annihilations is beset with large uncertainties relating
to unknown diffusion parameters combined with uncertainties in the
halo distribution. As an example, in \cite{salati} it is pointed out
that the estimated flux may vary by almost a factor 10 up or down,
for models that predict the measured B/C ratio (i.e., the relation
between secondary and primary cosmic rays) within the uncertainties.

Hence, it might seem that the models producing too many antiprotons
in Fig.~\ref{fig:deboer-pbar} are not really excluded if the
uncertainties of the antiproton propagation are included. However,
the results of a factor of around 50 uncertainty in \cite{salati}
are only valid for a relatively smooth halo profile, so we need to
investigate this in some more detail for the de Boer et al.\
profile. To do that, we will recalculate the antiproton fluxes with
the propagation code in \cite{salati} and with propagation
parameters chosen to be as extreme as we can allow from other cosmic
ray data (mainly B/C).

The main reason for the large uncertainties found in \cite{salati}
is a degeneracy (for the secondary signal) between the height of the
diffusion box and the diffusion parameter. If we increase the height
of the diffusion box, we would get a larger secondary signal because
cosmic rays can propagate longer in the diffusion box before
escaping. This can be counterbalanced by increasing the diffusion
coefficient to make the cosmic rays diffuse away faster from the
galactic disc. Hence, for the secondary signal, which originates in
the galactic disc, we can get acceptable fits by changing these
parameter. For the dark matter signal though, the effect of these
changes is different. If we increase the height of the diffusion
box, we also increase the volume in which annihilations occur, and
the total flux increases more than is counterbalanced by an increase
in the diffusion coefficient. In \cite{salati} they then found that
for a NFW (or isothermal) profile, the different acceptable
configurations of diffusion height and diffusion coefficient would
cause an increase/decrease of up to a factor of 10 compared to the
median value.

%
\begin{table}
\begin{center}
{\begin{tabular}{@{}cccccc@{}}
\hline
\hline
{\rm case} &  $\delta$  & $K_0$                 & $L$   & $V_c$    & $V_{a}$ \\
           &            & [${\rm kpc^{2}/Myr}$] & [kpc] & [km/sec] & [km/sec] \\
\hline
{\rm max} &  0.46  & 0.0765 & 15 & 5    & 117.6 \\
{\rm med} &  0.70  & 0.0112 & 4  & 12   & 52.9  \\
{\rm min} &  0.85  & 0.0016 & 1  & 13.5 & 22.4  \\
\hline
\hline
\end{tabular}}
\end{center}
\caption{ Astrophysical parameters of the cosmic ray galactic
propagation models giving the maximal, median and minimal primary
antiproton fluxes compatible with B/C analysis~\cite{salati}.
$\delta$ is the exponent for the diffusion coefficient's rigidity
dependence, $K_0$ is the normalization of the diffusion coefficient,
$L$ is the half-height of the diffusion box, $V_c$ is the galactic
wind and $V_a$ is the reacceleration velocity (see \cite{salati} for
more details). \label{tab:propagation}}
\end{table}
%

Let us now focus on the de Boer et al.\ profile. The most important
parts for the antiproton signal are the rings that decay
exponentially with the height above the galactic plane. Hence, this
source distribution is much more concentrated to the galactic disc
than a smooth halo profile is. We therefore expect a more modest
change of the fluxes with the extreme values of the diffusion
parameters than found in \cite{salati}.
As a complement to our calculations with DarkSUSY, we have derived
the antiproton fluxes in the de Boer model for the extreme diffusion
parameters given in \cite{salati} and which we recall in
Table~\ref{tab:propagation}. This calculation has been done with the
propagation code in \cite{salati}.

\begin{figure}[t!]
{\centerline{\includegraphics[width=0.8\textwidth]{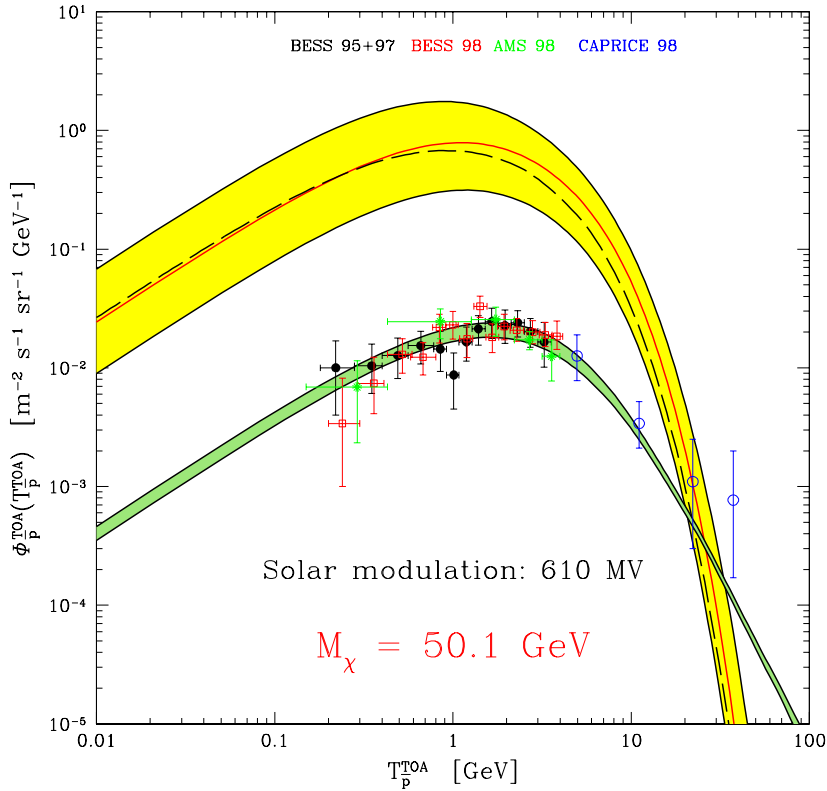}}}
\vskip -0.3cm
 \caption{ A supersymmetric model that provides a good
fit to the EGRET data has been selected and its antiproton yield has
been carefully derived. It is featured by the red solid line in the
case of the median cosmic ray configuration of
table~\ref{tab:propagation}. Predictions spread over the yellow band
as the cosmic ray propagation parameters are varied from the minimal
to maximal configurations. The long-dashed black curve is calculated
with DarkSUSY for a standard set of propagation parameters (see
\cite{msugra-indirect} for details).
The narrow green band stands for the secondary component.
As is evident from this figure, the antiproton fluxes for this
example model clearly overshoots the data.}
\label{fig:pbar_prim_vs_sec}
\end{figure}

For illustration, we show in Fig.~\ref{fig:pbar_prim_vs_sec} the
full yield of primary antiprotons for a selected supersymmetric
configuration for which the agreement with the EGRET data is very
good -- $\chi^{2}$ of order 3 (neutralino mass of 50.1 GeV and
derived boost factor of 69).
%
For the sake of completeness, we have taken into account tertiary
antiprotons as in~\cite{lars_joakim_ter} and computed the effect of
diffusive reacceleration with the help of the same Crank-Nicholson
scheme as in~\cite{donato_2001}. Solar modulation has been modeled
with the simple force--field approximation where a Fisk parameter of
$\phi_F = 610$ MV has been assumed.
The red solid curve in Fig.~\ref{fig:pbar_prim_vs_sec} corresponds
to the median cosmic ray configuration of
table~\ref{tab:propagation}. The yellow band is delimited by the
extreme configurations and gives an indication on how well the flux
of neutralino induced antiprotons can be derived in the case of the
de Boer et al.\ dark matter distribution.
For the maximal cosmic ray model, we observe an increase in the
antiproton flux by a factor of 2.5 whereas for the minimal
configuration, we find a decrease of a factor of 2.6 -- both at
energies lying between 0.4 and 0.6 GeV. The total width corresponds
therefore to an overall factor of $\sim 6.5$ to be compared to a
factor of $\sim 50$ in the case of an NFW dark matter halo.
As expected, the uncertainties are much smaller in the de Boer et al
model compared to more conventional halo profile. Again, the main
reason for this is that the dark matter in the de Boer model is
located much closer to the galactic plane and its distribution is
reminiscent of the gaseous disc of the Milky Way that is responsible
for the conventional secondary background. The latter is featured in
Fig.~\ref{fig:pbar_prim_vs_sec} as the narrow green band that has
been derived in~\cite{donato_2001} from the observed B/C ratio.
From this calculation we also get an estimate of the possible
uncertainties arising from different propagation codes. In
Fig.~\ref{fig:pbar_prim_vs_sec}, the long--dashed black curve has
been computed with the DarkSUSY package using a typical cosmic ray
propagation (as described in \cite{msugra-indirect}). Because
diffusive reacceleration has not been implemented in that case, the
flux falls more steeply close the neutralino mass. Notice however
that it is fairly similar to the red solid line, calculate with the
propagation code in \cite{salati}. We can therefore conclude that
diffusive reacceleration does not substantially modify the shape of
the antiproton spectrum -- especially in the energy range of
interest. We also note that the two different propagation codes,
DarkSUSY and the code of \cite{salati}, gives very good agreement
and should not constitute any additional theoretical error.
%
Notice finally that the primary yellow uncertainty band is at least
an order of magnitude above the secondary green component. The
latter alone can already account for the antiproton observations as
is clear in Fig.~\ref{fig:pbar_prim_vs_sec}. We conclude that the
supersymmetric model that has been selected here because it provides
a good fit to the EGRET excess should be excluded as it by far
overproduces antiprotons.
In Fig.~\ref{fig:pbar_prim_vs_sec} we also compare with the BESS
data at other energies than our previously selected energy bin. It
should be evident from the figure that the model overproduces
antiprotons at essentially every measured energy.

We have here compared with one example model, but this argument can
be made more general. In Fig.~\ref{fig:deboer-pbar}b we show with a
solid line how far down we could shift the models down by going to
the extreme minimal model. As can be seen, the antiprotons are still
overproduced by a factor of 2--10 for the models with good fits to
EGRET data. It is therefore difficult to see how the dark matter
interpretation of the EGRET data could be compatible with the
antiproton measurements.

\subsection{Other antiproton uncertainties}
\label{sec:pbar-uncert2}

Above we mentioned the antiproton uncertainties coming from our lack
of knowledge of the propagation model. We also have uncertainties
arising from the assumptions in the approach of de Boer et al. E.g.,
the boost factor is assumed to be independent of position in the
Galaxy. This is probably not a very well-justified assumption as we
would expect the boost factor to depend on the formation history of
that particular region in the Galaxy. If the boost factors do depend
on the position in the Galaxy, then the boost factors for the flux
at \emph{reception} at Earth can be different for gamma rays and
antiprotons since the signal in general come from different parts of
the Galaxy (the gamma signal can e.g.\ be dominated by annihilation
at the galactic center, whereas the antiprotons could come from a
more nearby structure). We will here investigate what the
uncertainties on the boost of the antiproton flux \emph{as seen at
the Earth} could be if we relax the assumption of space-independent
boost factors.

In particular, galactic tidal interactions should destroy
clumps\cite{clumps} in the inner (older) parts of the Milky Way and
especially in the inner ring, which is the most prominent feature in
the de Boer model when it comes to the fits to the gamma-ray fluxes.
On the other hand, the antiprotons that are detected at the Earth
originate from a broad region of the Milky Way halo. A significant
portion is produced in the inner ring, but a substantial part comes
also from the outskirts where clumps have survived. The correct
\emph{effective} boost factor which should have been implemented in
the antiproton calculations should therefore probably be larger than
the value derived from assuming a spatially constant boost factor.

We can quantify these uncertainties though. If we take away the
inner ring, the antiproton fluxes goes down a factor of 2.0. If we
take away the outer ring, the antiproton fluxes go down by a factor
of 1.4. And, finally, if we would take away the smooth triaxial
halo, the antiproton fluxes would go down by a factor of 2.2. Hence,
we can conclude that the antiproton fluxes at the Earth are
dominated by annihilation in the inner ring and the smooth triaxial
halo. Since the antiprotons coming from the inner ring originate in
the same place as the gamma rays that dominate the flux in region A,
the boost factor derived from the gamma flux should be very close to
the \emph{effective} boost factor for these antiprotons. As noted
above, this makes up half of the antiproton flux at Earth. For the
other half, coming mostly from the triaxial smooth halo, our use of
the same \emph{effective} boost factor as derived from gamma rays
from region A is most likely not correct. However, as noted above,
we would expect the boost factor to be larger in these less dense
regions, so our (and de Boer's) assumption of space independent
boost factors would lead to an underestimate of the antiproton flux.

Another argument -- slightly more technical -- points in the same
direction. The \emph{local} boost factor $B_{c}$ of a clump of mass
$M_{c}$ is defined~\cite{clump} with respect to some value of
reference $\rho_{0}$ through the relation
\begin{equation}
{\displaystyle \int}_{\rm clump} \!\!\!\!\! d^{3} \vec{x} \;\;
\delta \rho^{2} \left( \vec{x} \right) \; = \; \rho_{0} \, M_{c} \, B_{c} \;\; ,
\end{equation}
where $\delta \rho$ is the dark matter density inside the
substructure. If large boost factors are needed to explain the EGRET
data, this means that the dark matter is tightly packed even inside
the inner ring. That region is already fairly dense with a density
of reference $\rho_{0} = 4.5$ GeV cm$^{-3}$. On the contrary,
antiprotons originate from a larger domain where the neutralino
density is certainly smaller on average than inside the inner ring.
If the clumps have basically the same mass and density profile all
over the Milky Way halo -- which is probably correct in the case
where very small substructures dominate, as they should have
survived the galactic tides -- the product $\rho_{0} \, B_{c}$
remains constant. We are forced to the conclusion that the actual
\emph{effective} boost value which should be used for antiprotons is
once again larger than in the case of the EGRET data. Therefore
assuming identical \emph{effective} boost factors (as seen at Earth)
for the photon and antiproton flux is very conservative. The lower
antiproton limit featured in Fig.~\ref{fig:deboer-pbar}b should
actually be shifted upwards, even further above the BESS 98 stripes.

\section{Comparison with direct detection}

\begin{figure}[t]
\centerline{\epsfig{file=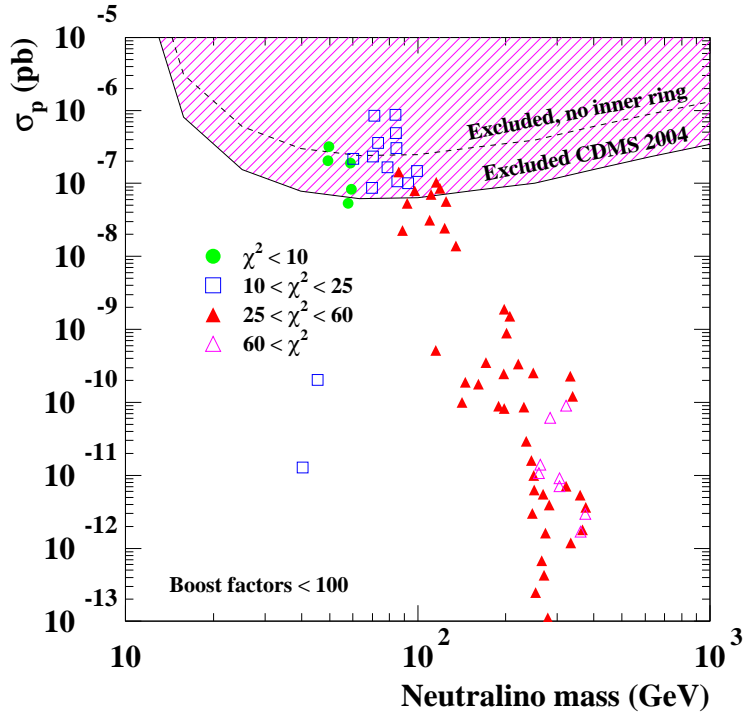,width=0.75\textwidth}}
\caption{The spin-independent cross section for direct detection
versus mass is shown for models with boost factors below 100. The
$\chi^2$ of the fits has been indicated with different symbols. The
CDMS limits \cite{cdms} have been lowered by a factor of $\rho_{\rm
Boer}/\rho_0 = 1.93/0.3$ as the density in the de Boer is higher
than assumed in the analysis of \cite{cdms}. Also indicated is the
limit in case the inner ring would not contribute to the direct
detection rates (reduced by factor of $0.5/0.3$ compared to
\cite{cdms}.} \label{fig:deboer-dir}
\end{figure}

With the enhanced dark matter densities in the de Boer et al.\ model
one may wonder if these models wouldn't have been seen already in
direct detection experiments. To make a simple estimate of this, we
plot the scattering cross section on protons versus mass in
Fig.~\ref{fig:deboer-dir}. We compare with the latest CDMS results
\cite{cdms}, but since those results apply for a local dark matter
density of $0.3$ GeV/cm$^3$, we have rescaled the limits with the
local density in the de Boer et al.\ model. To be more specific, we
have lowered the limits by $\rho_{\rm Boer}/\rho_0 = 1.93/0.3$,
which is the ratio of the local dark matter density in the de Boer
et al.\ model and a `standard' halo model. As can be seen, most
models would seem to be excluded by the latest CDMS results
\cite{cdms}. However, in deriving these limits, a standard gaussian
velocity distribution has been assumed.  This is most probably a
reasonable approximation for the triaxial halo in the de Boer et
al.\ model, but most of the local density comes from the inner ring,
where not much is known about the velocity. If the inner ring would
rotate along with the Galaxy, we expect lower rates than indicated
in the figure since the relative velocity between the dark matter
and us would be lower. However, if it counter-rotates, we would
expect even higher rates. If we would be conservative and assume
that the inner ring would produce no significant rates in direct
detection experiments, the limits from CDMS would be reduced by a
factor of $0.5/0.3$ (as the local density from the smooth halo is
0.5 GeV/cm$^3$ in the de Boer et al.\ model). In
Fig.~\ref{fig:deboer-dir} we indicate this exclusion limit with a
dashed line. As can be seen, some of the models of them would then
fall below the exclusion limit, but some would still be above.

Hence, one can conclude that the de Boer et al.\ models produce
significant rates in direct detection experiments, and some of the
good models are already excluded even if uncertainties of the
velocity distributions are taken into account. However, we cannot
exclude all of the good models, due to these uncertainties, but many
of the models giving the best fits are well within reach with future
improved direct detection experiments.

\section{Discussion}

We have here investigated the idea put forward by de Boer et al.\
that the observed EGRET excess of diffuse gamma rays could be due to
dark matter annihilation in the Milky Way. This is an interesting
idea, but as far as we can see the model does not seem very
plausible when other constraints are taken into account. The
strongest constraint is probably the antiproton flux, which would be
overproduced far above the flux measured by BESS. Even if the
uncertainties of the antiproton flux are included, the antiproton
flux is still more than a factor of 2--10 above the BESS
measurements. From the analysis in this paper one also find (see,
Fig.~\ref{fig:deboer-pbar}b) that models that would be compatible
with the antiproton data would always have very bad fits to the
EGRET data (reduced $\chi^2$ $>$ 60/6). Actually, the optimized
background model in \cite{mosk-strong-reimer} would produce a much
better fit (reduced $\chi^2$ $\sim$ 22/6 ; which can also be read of
at the high-mass end in Fig.~\ref{fig:deboer-boost}b, where the
signal does not contribute much to the fits). Hence, the models with
acceptable antiproton fluxes are not very interesting as they give
worse fits to the gamma excess than the more conventional models.

Other objections to the model come from astrophysics. E.g., the
density in the disc in the de Boer model is a little bit too high to
be compatible with stellar motions in the solar neighborhood. This
could of course be circumvented by adjusting the densities in the de
Boer et al.\ model to have a dip at our location in the solar
system. Such fine tunings are not very appealing though, especially
as the model already as it stands has a local minimum at the solar
neighborhood. One has to be careful interpreting these results on
the disc surface density though as there are large uncertainties and
model dependencies. To be conservative, we interpret these too high
disc surface mass densities as a hint of problems with the model.
The actual density profiles of the rings (exponential fall-off away
from the disc) is also not what you expect from dark matter. In
fact, we believe that the de Boer et al.\ model is most likely a fit
of the \emph{baryonic} matter distribution of the Galaxy, and not
the dark matter density.
One could also wonder if the background models could be improved
further and provide a even better fit to the EGRET data. In fact,
the optimized background model in \cite{mosk-strong-reimer} already
gives a good fit to the EGRET data -- especially when
conventional/larger uncertainties of \hbox{$\sim$ 15\%} for EGRET's
observed gamma fluxes are adopted (reduced $\chi^2$ $\sim$ 5/6).
This optimized model was accomplished by tuning the electron and
proton injection spectrum without violating any of the cosmic ray
constraints, such as B/C.
Hence, there is no real need for dark matter to explain the EGRET
data.
New experiments are on their way though with GLAST flying in 2007
and with new and better data, we will hopefully get a better
understanding of the gamma-ray sky and if there are any (clear)
signals of dark matter out there.

\bigskip
M.G.~would like to thank C.~Flynn and J.~Holmberg for useful
discussions regarding the local galactic surface density. L.B.~and
J.E want to thank I.~Moskalenko and W.~de Boer for useful
discussions and are grateful to the Swedish Science Research Council
(VR) for support. J.E.~also thanks A.~Strong and O.~Reimer for
discussions.
P.S.~acknowledges a support from the French programme national de
cosmologie PNC.


\end{document}